\documentclass[12pt,a4paper,twoside]{article}

\newcommand{\be}{\begin{equation}}
\newcommand{\ee}{\end{equation}}
\newcommand{\beqs}{\begin{eqnarray}}
\newcommand{\eeqs}{\end{eqnarray}}

\newcommand{\Tr}{{\rm Tr}}

\newcommand{\D}{{\rm D}}

\expandafter\ifx\csname mathbbm\endcsname\relax

\else

\fi
\begin{document}
\begin{titlepage}
\begin{flushleft}  
       \hfill                       RU-02-4-B\\
       \hfill                       June 2002\\
\end{flushleft}
\vspace*{3mm}
\begin{center}
{\LARGE  Seiberg-Witten map and topology\\}
\vspace*{12mm}
\large Alexios P. Polychronakos\footnote{E-mail: poly@teorfys.uu.se} \\
\vspace*{5mm}
{\em Physics Department, Rockefeller University \\
New York, NY 10021, USA \/}\\
\vspace*{4mm}
and\\
\vspace*{4mm}
{\em Physics Department, University of Ioannina \\
45110 Ioannina, Greece\/}\\
\vspace*{15mm}
\end{center}

\begin{abstract}
The mapping of topologically nontrivial gauge transformations in noncommutative
gauge theory to corresponding commutative ones is investigated via the operator
form of the Seiberg-Witten map. The role of the gauge transformation
part of the map is analyzed. Chern-Simons actions are examined
and the correspondence to their commutative counterparts is clarified.

\end{abstract}

\end{titlepage}

\section{Introduction}

The Seiberg-Witten map \cite{SW} establishes a correspondence between 
noncommutative gauge potentials for different values of the noncommutativity parameter.
It has the remarkable and essentially defining property that a noncommutative
gauge transformation in one field results in a gauge transformation in the mapped
field, therefore preserving gauge-invariant observables. This map can, in principle,
be used to map a noncommutative gauge theory to a commutative one, although
the action will in general become nonlocal. As such, it has been extensively 
studied \cite{JSW}-\cite{KiSo}

The differential form of the map, relating gauge theories at infinitesimally
differing noncommutativity parameters $\theta$, defines the Seiberg-Witten equation.
This equation has a non-covariant form.
Important steps in the identification of its solutions were done by Liu \cite{Liu} and 
especially by Okawa and Ooguri \cite{OO}, by identifying the gauge-invariant 
abelian commutative field strength produced by the map at $\theta=0$.
(Similar results were also obtained in \cite{MS}, \cite{LM}.)
The response of this map under nontrivial gauge transformations, however, remains
obscure. As an example, it was shown by Grandi and Silva \cite{GS} that the 
3-dimensional Chern-Simons
action remains invariant under the map. On the other hand, noncommutative
gauge theory in odd dimensions exhibits topologically nontrivial gauge transformations
which imply a level quantization of the Chern-Simons action even in the U(1) case
\cite{NP,BLP}. 
In \cite{NP}, in particular, such a transformation was explicitly demonstrated. Since
commutative U(1) transformations are trivial, the Seiberg-Witten map must fail
for such cases.

Noncommutative gauge theory is usually formulated in terms of star-products. 
A different and quite effective formulation is the operator language \cite{GN,AW},
in which covariant derivatives in the noncommutative directions become 
operators acting on a Heisenberg-like Hilbert space. Gauge 
transformations simply become unitary conjugations of these operators.
Any operator written entirely in terms of covariant derivatives is explicitly gauge
covariant. Yang-Mills and Chern-Simons actions can be compactly written in this 
language in a universal way for any U(N) gauge group \cite{APA,APB}.

The Seiberg-Witten map is usually written in terms of ordinary commutative functions in
the star-product formulation. It is more convenient for our purposes to use the operator
form of the map. Topologically nontrivial gauge configurations
(solitons) or gauge transformations are easily written in the operator language
and are, thus, amenable to explicit analysis.

\section{The operator map}

The operator form of the Seiberg-Witten map for a fully noncommutative space
was derived by Kraus and Shigemori \cite{KS}.
In the form presented there, however, it induced a non-unitary gauge transformation
spoiling the hermiticity properties of the fields.
For completeness, we give here a derivation suited to our purposes.

We consider a $D$-dimensional space with $2n$ noncommutative coordinates 
satisfying
\be
[ x^\alpha , x^\beta ] = i \theta^{\alpha \beta}
\ee
and $D-2n$ commutative ones. We will use middle greek indices for the full space
($\mu,\nu,\dots = 1,\dots D$), early greek indices for the purely noncommutative
dimensions ($\alpha,\beta,\dots = 1, \dots 2n$) and latin indices for the commutative
dimensions ($i,j,\dots = 2n+1,\dots D$). We consider $U(N)$ gauge theory, 
in which case the gauge fields $A_\mu$ are $N \times N$ hermitian matrices.
The Seiberg-Witten equation for the change of the components of the gauge field 
$A_\mu$ under a small change in the (constant, c-number) noncommutativity 
tensor $\theta^{\alpha \beta}$ reads
\be
\delta A_\mu = -\frac{1}{4} \delta^{\alpha \beta} \{ A_\alpha , 
\partial_\beta A_\mu
+ F_{\beta \mu} \} \label{SW}\ee
In the above, $\{ \, , \, \}$ denotes anticommutator and all matrix multiplications 
are understood to involve $*$-products of the matrix elements 
defined in terms of $\theta^{\alpha \beta}$. with the field strength defined as
\be
F_{\mu \nu} = \partial_\mu A_\nu - \partial_\nu A_\mu - i [ A_\mu , A_\nu ]
\ee
From (\ref{SW}) the change of $F_{\mu \nu}$ and of the gauge transformation
parameter $\lambda$ can be deduced.

In the operator formulation, we work with the covariant derivative operators
$D_\mu = i \partial_\mu + A_\mu$. For the commutative dimensions, $D_i$ is
a bona-fide differential operator. For the noncommutative dimensions, on the
other hand, derivatives can be realized via the adjoint action of coordinates
themselves. We define the operators
\be
i \partial_\alpha = \omega_{\alpha \beta} x^\beta
\ee
where $\omega_{\alpha \beta}$ is the two-form inverse to $\theta^{\alpha \beta}$:
\be
\omega_{\alpha \beta} \theta^{\beta \gamma} = \delta_{\alpha}^{\gamma}
\ee
The above operators satisfy 
\be
[i \partial_\alpha , x^\beta ] = i \delta_\alpha^\beta
\ee
and thus act as derivatives upon commutation. The noncommutative covariant 
derivatives $D_\alpha$ become operators acting on the same space as the $x^\alpha$:
\be
D_\alpha = i \partial_\alpha + A_\alpha = \omega_{\alpha \beta} x^\beta + A_\alpha
\ee
and can be considered as the fundamental dynamical objects. The separation of 
$D_\alpha$ into derivative and gauge potential is gauge-dependent and can be
changed by a gauge transformation, understood as a conjugation of all $D_\mu$
by an ($x^i$-dependent) unitary operator. Due to the nontrivial commutator of
$[ \partial_\alpha , \partial_\beta ]$ (viewed as ordinary operators rather than
through their adjoint action), the gauge field strength becomes
\be
i F_{\alpha \beta} = [ D_\alpha , D_\beta ] + i \omega_{\alpha \beta}
\ee

The operators $x^\alpha$ are, essentially, a set of $n$ canonical pairs. Their
irreducible representation consists of $n$ copies of the quantum mechanical
Heisenberg-Fock space. By considering, instead, the direct sum of $N$ 
copies of the irreducible representation we can include the $U(N)$ part in the
operator structure. Labeling the copies with an extra index $a=1,\dots N$,
$x^\alpha$ and $\partial_\alpha$ act trivially on $a$ while the operator $D_\alpha$
becomes an $N \times N$ operator matrix in this space. Thus, we can deal both 
with $U(1)$ and $U(N)$ gauge theories in the operator language without explicitly
modifying the formalism.

In deriving the response of $D_\mu$ under a change of $\theta^{\alpha \beta}$
it is important to realize that the operators $x^\alpha$ are, themselves, $\theta$-dependent
and they also respond to the change. To make this explicit, consider the case 
of two noncommutative dimensions $x^{1,2}$ with $\theta^{12} = \theta$. The coordinates
can be realized in terms of a canonical pair $[q^1, q^2] = i$ as
\be
x^\alpha = \sqrt{\theta} q^\alpha 
\ee
The mapping
from an operator $\hat f$ to a commutative function $f$, such as the gauge fields
appearing in (\ref{SW}), is through the Weyl ordering procedure, written explicitly
in terms of the Fourier transform of $f$, ${\tilde f} (k)$
\be
{\hat f} = \int d^2 k  {\tilde f} (k) e^{i {\vec k} \cdot {\vec x}} =
\int d^2 k  {\tilde f} (k) e^{i \sqrt{\theta} {\vec k} \cdot {\vec q}}
\ee
For a small variation of $\theta$ we have
\begin{eqnarray}
\delta {\hat f} &=& \int d^2 k  \delta {\tilde f} (k) e^{i \sqrt{\theta} 
{\vec k} \cdot {\vec q}} + \delta \theta
\frac{i}{2\sqrt{\theta}} \int d^2 k  {\tilde f} (k) {\vec k} \cdot {\vec q} 
e^{i \sqrt{\theta} {\vec k} \cdot {\vec q}} \cr
&=& \int d^2 k  \delta {\tilde f} (k) e^{i \sqrt{\theta} {\vec k} \cdot {\vec q}} +
\delta \theta \frac{i}{2\theta} \epsilon_{\alpha \beta} 
q^\alpha [ q^\beta , {\hat f} ]
\end{eqnarray}
In the above, the first term is due to the change of the commutative
function $f$ itself, and will be denoted $\delta_f {\hat f}$, while the second
term is due to the change of the coordinates. This can easily be generalized
to $2n$ dimensions as
\be
\delta {\hat f} = \delta_f {\hat f} +
\frac{i}{2} \delta \omega_{\alpha \beta} x^\alpha [ x^\beta , {\hat f} ]
\label{dtotal}\ee
expressed in terms of the variation $\delta \omega_{\alpha \beta} =
\omega_{\alpha \gamma} \omega_{\beta \delta} \delta \theta^{\gamma \delta}$.
The Seiberg-Witten transformation expresses only the variation of the commutative
functions $A_\mu$; to find the variation of the operators $A_\mu$ or $D_\mu$
the second term above must be included.

We now have all the ingredients. Expressing $\partial_\mu A_\nu$ and 
$F_{\mu \nu}$ in (\ref{SW}) in terms of their operator expressions and taking 
into account (\ref{dtotal}) we obtain, after some algebra,
\be
\delta D_\mu = - \omega_{\mu \nu} \delta \theta^{\nu \rho} D_\rho +
\frac{i}{4} \delta \theta^{\alpha \beta} \{ D_\alpha , [ D_\beta , D_\mu ] \}
+ i [ \delta G , D_\mu ]
\label{OSW}
\ee
where the infinitesimal operator $\delta G$ is given by
\be
\delta G = \frac{1}{4} \delta \theta^{\alpha \beta} \{ i\partial_\alpha , D_\beta \}
\label{dG}
\ee

We observe that the above transformation contains a covariant piece,
involving only $D_\mu$, plus a non-covariant piece involving also $\partial_\alpha$.
This last piece, however, amounts to a gauge transformation. We can redefine the
map to include any gauge transformation we want, and therefore we may discard
that piece, to obtain an explicitly covariant form of the equation. In fact, it is somewhat
neater to express it in terms of the covariant coordinate operators
\be
X^\alpha = \theta^{\alpha \beta} D_\beta
\ee
and the remaining covariant derivatives in the commutative directions $D_j$.
We obtain
\be
\delta X^\gamma = \frac{i}{4} \delta \omega_{\alpha \beta} \{ X^\alpha , 
[ X^\beta , X^\gamma ] \} 
\label{CSWX}
\ee
\be
\delta D_j = \frac{i}{4} \delta \omega_{\alpha \beta} \{ X^\alpha , [ X^\beta , D_j ] \}
\label{CSWD}
\ee
These have a very suggestive form. If instead of $X^\alpha , X^\beta$ 
in the above
we substitute $x^\alpha , x^\beta$ then (\ref{CSWX},\ref{CSWD}) become
identical to the second term in (\ref{dtotal}). The above equations, therefore, are
a covariant version of the change of the corresponding operators due to the change
of the `scale' of its underlying space variables. 

The above formula for $X^\gamma$ can also be rewritten in the form
\be
\delta X^\gamma = \frac{1}{4} \delta \theta^{\alpha \beta} \left[
\{ F_{\alpha \beta}, X^\gamma \} + 2 \omega_{\alpha \beta} X^\gamma
- 2i D_\alpha X^\gamma D_\beta \right]
\ee
This also provides an immediate
generalization of the operator transformations for any fields that transform in the adjoint,
fundamental or antifundamental representation of the gauge group. On such 
fields,  denoted $\rm A$, $\rm f$ and $\bar {\rm f}$, the gauge transformation $U$ 
acts on both sides, to the left only, or to the right only, respectively. Therefore, 
we simply substitute the covariant $X^\alpha$ or $D_\alpha$ by ordinary 
$x^\alpha$ and $\partial_\alpha$ on the side of the operator field where $U$ 
does not act. We have
\be
\delta {\rm A} = \frac{1}{4} \delta \theta^{\alpha \beta} \left[
\{ F_{\alpha \beta} , {\rm A} \} + 2 \omega_{\alpha \beta} {\rm A} 
-2i D_\alpha \, {\rm A} \, D_\beta \right]
\ee
\be
\delta {\rm f} = \frac{1}{4} \delta \theta^{\alpha \beta} \left[
F_{\alpha \beta}  {\rm f}  + 2 \omega_{\alpha \beta} {\rm f}
+ 2 D_\alpha \, {\rm f} \, \partial_\beta \right]
\ee
\be
\delta {\bar {\rm f}} = \frac{1}{4} \delta \theta^{\alpha \beta} \left[
{\bar {\rm f}} F_{\alpha \beta} + 2 \omega_{\alpha \beta} {\bar {\rm f}}
+ 2 \partial_\alpha \, {\bar {\rm f}} \, D_\beta \right]
\ee

\section{The commutative limit and the role of the gauge transformation}

The gauge variation connecting the covariant equations (\ref{CSWX},\ref{CSWD})
to the original Seiberg-Witten equations assumes the form
\be
\delta G = \frac{1}{4} \delta \omega_{\alpha \beta} \{ x^\alpha , X^\beta \}
\label{dGX}
\ee
We observe that for $X^\alpha = x^\alpha$, the variation $\delta G$ vanishes and thus
for $X^\alpha$ close to $x^\alpha$ the above transformation will remain small. As
$X^\alpha$ becomes increasingly different than $x^\alpha$, however, $\delta G$
becomes more important. Its action is to tend to `realign' $X^\mu$ and $x^\mu$ as
close as possible.

This is relevant to the commutative limit. For every nonzero $\theta^{\alpha \beta}$
the mapping from $\theta$ to $\theta+\delta \theta$ is smooth. In fact, for the covariant
equations, operator gauge transformations are mapped trivially without any 
dependence on the fields:
\be
U( \theta + \delta \theta ) = U (\theta )
\ee
in contrast to the original equations. There is no cocycle. The limit $\theta \to 0$,
on the other hand, could be singular. For the operators $X^\alpha$, or $D_\alpha$, to
correspond to smooth finite gauge potentials in this limit they must have the scaling
\be
X^\alpha = x^\alpha + \theta^{\alpha \beta} A_\beta ( x ) + {\cal O} (\theta^2) ~,~~~ 
iD_\alpha = \omega_{\alpha \beta} x^\beta + A_\alpha ( x ) + {\cal O} (\theta )
\label{scale}
\ee
where $A_\mu (x)$ is a smooth function of the $x^\mu$ not explicitly depending on
$\theta$. To make the scaling more explicit, we use $\theta$-independent
Darboux operators (canonical pairs) $q_\alpha$ satisfying
\be
[ q_\alpha , q_\beta ] = i \epsilon_{\alpha \beta}
\ee
with $\epsilon_{2k-1,2k} = -\epsilon_{2k,2k-1} =1$, else zero, and write
\be
x^\alpha = {\sqrt \theta \,}^{\alpha \beta} q_\beta
\ee
where  the `square root of theta' matrix satisfies
\be
{\sqrt \theta \,}^{\alpha \gamma} {\sqrt \theta \,}^{\beta \delta} \epsilon_{\gamma \delta} 
= \theta^{\alpha \beta}
\ee
Then the scaling at the limit $\theta \to 0$ should be
\be
X^\alpha = {\sqrt \theta \,}^{\alpha \beta} q_\beta + \theta^{\alpha \beta}
A_\beta ( {\sqrt \theta \,} q ) + {\cal O} ({\sqrt \theta \,}^3 )
\label{scaleq}
\ee
The role of $\delta G$ is to ensure, bar other obstructions, that the above 
relations are satisfied. Indeed, we can check that the fully covariant 
equations (\ref{CSWX}) do not admit (\ref{scaleq}) as a solution to order
$\theta^{3/2}$, while the full equations (\ref{OSW}), including $\delta G$,
are satisfied by (\ref{scaleq}).

We are therefore led to defining a class of 
{\it admissible gauge transformations}
$\delta G$, to accompany the covariant equations (\ref{CSWD},\ref{CSWX}),
having the property that they admit (\ref{scaleq}) as a solution and therefore
leading to a smooth commutative limit. The original Seiberg-Witten gauge
(\ref{dG}) or (\ref{dGX}) is such a gauge, but it is by no means unique. 
We shall give an example of an alternative gauge in a subsequent section.

Alternatively, we could simply solve the covariant equations and a posteriori
`rehabilitate' the solution by performing an appropriate gauge transformation. If such a
transformation can be found which ensures that (\ref{scaleq}) is satisfied, we will
have a smooth commutative limit. Such a transformation, however, will in general
depend on the specific form of the field. Therefore, the mapping between gauge
transformations will involve a dependence on the field, recovering a cocycle
\cite{JSW,JP}.

The mapping to the commutative case is essentially a reduction. The 
noncommutative gauge groups are always isomorphic and the mapping is
trivial. In the commutative limit, however, we have a reduction similar to the
reduction of, say, $SO(3)$ to the planar Euclidean group. It is known that a 
specific choice of basis must be made in the representations of the original
group in order to obtain representations of the reduced group. The admissible
gauge transformation does just that.

We conclude by warning that an admissible gauge transformation admits
(\ref{scaleq}) as solutions but does not a priori {\it guarantee} that such a form
will be reached at the commutative limit. This is the source of the change of
topology between the noncommutative and commutative cases, and will be
the subject of the next section.

\section{Topologically nontrivial gauge transformations}

The Chern-Simons action is invariant under an infinitesimal Seiberg-Witten
transformation \cite{GS}. If the transformation worked all the way to the limit 
$\theta \to 0$, then we would obtain a mapping between the commutative 
and noncommutative theory preserving the action. 

On the other hand, odd-dimensional noncommutative gauge theory exhibits
topologically nontrivial gauge transformations, even in the U(1) case, due 
to which the coefficient of the Chern-Simons action must be quantized \cite{NP}.
Since there are no such nontrivial transformations in the commutative U(1)
case, the mapping between the theories must fail.

In this section we shall examine the Seiberg-Witten mapping of topologically
nontrivial gauge transformations and demonstrate their singularity as $\theta$
vanishes, thereby resolving the issue.

We consider a noncommutative plane with a commutative third direction,
that is, $D=3$, $n=1$, $\theta^{\alpha \beta} = \theta \epsilon^{\alpha \beta}$.
The topology of gauge transformations is classified by the group
\be
\pi_1 \left( U(N) \right) = Z
\ee
where $N$ is any integer, corresponding to a finite truncation (regularization)
of the Hilbert space of the Heisenberg algebra defined by $x^1, x^2$. Gauge
transformations are classified according to their winding number $w \in Z$.
The `prototype' gauge transformation with winding number $w=1$ is
\be
U_o = 1 + \left( e^{i \phi_o (t) } -1 \right) |0\rangle \langle 0 | ~,~~~
\phi_o (+\infty ) - \phi_o (-\infty ) = 2\pi
\ee
where $t = x^3$ and $|0\rangle$ is any state in the Hilbert space, taken above
to be the `vacuum' of the oscillator operator $x^1 + i x^2$. $U$ is unitary and
well-defined for any $\theta$, although it maps to a singular commutative field
in the limit $\theta \to 0$.

We shall take the gauge field to be a nontrivial gauge transformation of the 
trivial vacuum $A_\mu = 0$, that is,
\be
D_\mu = U^{-1}_o i \partial_\mu U_o
\ee
and shall evaluate the Seiberg-Witten mapping of the above field for varying
$\theta$. This will remain a ($\theta$-dependent) gauge transformation 
of the vacuum for all values of $\theta$. Indeed, if $D_\mu (\theta)$ is a solution
of the Seiberg-Witten transformation, 
then so is $U(\theta) D_\mu (\theta) U(\theta)$ provided that $U(\theta)$ satisfies
\be
i U^{-1} \delta U= \delta G (U^{-1} D U) - \delta G(D)
\label{UdG}\ee
Here $\delta G (D)$ is either the standard Seiberg-Witten gauge transformation
(\ref{dG}) or any other admissible gauge transformation. The trivial vacuum
$A_\mu = 0$ is always a solution of the equations (with $\delta G =0$).
Therefore if $U(\theta)$ satisfies (\ref{UdG}) above and the initial condition
$U(\theta_o ) = U_o$, then it will provide a solution of the equations mapping
$U_o$ to $U(\theta)$.

It is convenient to switch to oscillator variables. Define
\be
a = \frac{x^1 + i x^2}{\sqrt{2\theta}} ~,
~~~ [ a, a^\dagger ] = 1
\ee
The corresponding covariant coordinates are
\be
Z = \frac{X^1 + i X^2}{\sqrt{2\theta}} = U^{-1} a U
\ee
The gauge transformation (\ref{dGX}) in this case becomes
\be
i \delta G = \frac{i}{4} \delta \omega_{\alpha \beta} \{ x^\alpha , X^\beta \}
= \frac{\delta \theta}{4 \theta} \left( \{ a^\dagger , Z\} - \{ a , Z^\dagger \} \right)
\label{dGO}
\ee
Instead, we shall use the alternative gauge
\be
i \delta G' = i \delta G + \frac{\delta \theta}{4 \theta } 
\left( [ x^1 , X^1 ] + [ x^2 , X^2 ] \right)
= \frac{\delta \theta}{2 \theta } ( a^\dagger Z - Z^\dagger a )
\ee
It is important to stress that the above modified transformation is still admissible, 
as can be directly verified by the fact that it admits (\ref{scaleq}) as a solution.
This is simpler than (\ref{dGO}) and we shall use it for our calculation.

Due to the rotational symmetry of the vacuum configuration and of the original
transformation $U_o$, we shall choose $U(\theta)$ to also be rotationally
symmetric; that is,
\be
U(\theta) = \sum_{n=0}^\infty e^{i \phi_n} |n\rangle \langle n |
\ee
where $\phi_n$ are $\theta$-dependent phases and $|n \rangle$ are $a$-oscillator
states. Equation (\ref{UdG}) for $U$, then, with $\delta G'$ above as the gauge
transformation, becomes
\be
\delta \phi_n = \frac{\delta \theta}{\theta} \, n \, \sin (\phi_n - \phi_{n-1} )
\ee
In the limit $\theta \to 0$, $\phi_n$ and $\phi_{n-1}$ become almost equal
and thus we can approximate $\sin (\phi_n - \phi_{n-1} ) \simeq \partial \phi_n
/ \partial n$. The above equation in that limit becomes
\be
\left( \frac{\partial}{\partial \ln \theta} - \frac{\partial}{\partial \ln n} \right) \phi_n = 0
\ee
Defining the (positive) variable $r = \sqrt{2n \theta}$, the above equation admits
as solution
\be
\phi = \phi (r)
\ee
Since $r^2$ are the eigenvalues of the operator $2\theta a^\dagger a =
(x^1)^2 + (x^2)^2 + \theta$, the above simply expresses the fact that in the commutative
limit $\phi$ becomes a smooth function of the radial coordinate, thus reconfirming
the admissibility of the chosen gauge and the rotational symmetry of the configuration.

To derive the shape of $\phi(r)$ as $\theta \to 0$, we first note that $\delta \phi_0 = 0$,
and thus $\phi (r = 0) = \phi_o (t)$. Further, we note that the values $\phi_n =0,\pi$ 
are fixed points at which $\delta \phi_n = 0$. Since the initial condition is $\phi_0 = 
\phi_o$, $\phi_n = 0$ ($n>0$), we conclude that $\phi (r=\infty) = 0$. So $\phi (r)$
will interpolate between $\phi_o$ at the origin and $0$ at infinity, and the question
is which way.

To decide that, note that for $\phi_o = \pi - \epsilon$, $\sin (\phi_1 - \phi_0 ) <0$
and thus, for $\delta \theta / \theta <0$, $\phi_1$ will increase. The rest of the 
$\phi_n$ will follow suite, and thus the interpolation will be in the interval
$[ \pi -\epsilon , 0 ]$. Conversely, for $\phi_o = \pi + \epsilon \equiv -\pi + \epsilon$, 
$\sin (\phi_1 - \phi_0 ) >0$ and thus $\phi_1$ and the rest of $\phi_n$ will
decrease. In this case the interpolation will be in the interval $[ -\pi +\epsilon , 0]$.
For $\theta_o = \pi$, $\phi_n$ will remain $0$ (for $n>1$) and thus $\phi(r)$
will become a single spike at the origin. The spatial spread 
of the function $\phi(r)$ will keep diminishing with decreasing $\theta$, approaching
a zero-support function at  $\theta=0$.

We have concluded that the topologically nontrivial gauge transformation
is mapped by the Seiberg-Witten map to a {\it singular} abelian commutative
transformation. The singularity appears as a discontinuity in time. 
The spatial profile is smooth for all times $t<t_o$. 
As we approach the time $t = t_o$, in which $\phi_o (t_o ) = \pi$, 
the gauge transformation shrinks and at $t=t_o-\epsilon$ it becomes a spike at 
$r=0$ with amplitude $\pi$. At $t=t_o +\epsilon$ the amplitude of the spike becomes
$-\pi$ (which is gauge-equivalent to $\pi$) and for $t>t_o$ we have a smooth
profile again but with opposite sign. The transformation is singular,
producing
infinite derivatives and a singular gauge field. Such a gauge 
transformation is inadmissible in the commutative theory. Thus, gauge
field configurations which would be related to (smooth) commutative fields by a
topologically nontrivial gauge transformation become singular and decouple from
the theory, leaving only trivial topology.

The fact that the Seiberg-Witten map can be singular is well known. For a constant
magnetic field $B = \theta^{-1}$, for instance, the corresponding commutative field
is infinite \cite{SW}. What we have demonstrated above, however, is rather different: it is a
singularity which exists even for {\it zero} field strength and originates in the topology
of the gauge transformations. We expect it to persist in all odd-dimensional spaces.

\section{Chern-Simons invariants of the map}

We conclude this note by commenting on the invariance of Chern-Simons
actions under the Seiberg-Witten transformation. In the operator formulation the
$2n+1$-dimensional noncommutative Chern-Simons action is written in form notation
as \cite{APB}
\be
S_{2n+1}  =  \sum_{k=0}^n 
{n+1 \choose k+1} \frac{k+1}{2k+1}  (-\omega)^{n-k} \, \Tr \D^{2k+1}
\label{ShS}
\ee
where $\D = dx^\mu D_\mu$ is an operator one-form corresponding to the noncommutative
covariant derivative and $\omega$ is the two-form $\omega = \omega_{\alpha \beta} 
dx^\alpha dx^\beta$. A variation of the above action yields
\be
\delta S_{2n+1}  = \sum_{k=0}^n 
{n+1 \choose k+1} (k+1)  (-\omega)^{n-k} \, \Tr ( \D^{2k} \delta \D)
\ee
We may, now, use expression (\ref{OSW}) for the variation $\delta D$ and
calculate the variation $\delta S_{2n+1}$. The gauge transformation term $\delta G$
in (\ref{OSW}) produces no variation, since $S_{2n+1}$ is gauge invariant. 
Of the covariant terms in (\ref{OSW}), the first is linear in $D_\mu$ and preserves the
degree of any monomial in $D_\mu$, while the second involves three covariant 
derivatives and thus increases the degree of any monomial in $D_\mu$ by two. 
The variation of the highest-order term in (\ref{ShS}) $\Tr \D^{2n+1}$, then, 
under this transformation will be of order $2n+3$ and cannot be canceled by 
any lower-dimensional term. If $S_{2n+1}$ is to be invariant , this terms itself must
vanish. Se we examine the term
\begin{eqnarray}
\Tr ( \D^{2k} \delta \D) &=&\delta \theta^{\alpha \beta} \, \Tr 
\left( \{ D_\alpha , [ D_\beta , \D] \} \D^{2n} \right) \cr
&=& 2 \delta \theta^{\alpha \beta} \, \Tr \left( D_\alpha D_\beta \D^{2n+1} - D_\alpha 
\D D_\beta \D^{2n} \right)
\label{SWCS}
\end{eqnarray}
where we used cyclicity of trace. To analyze the above further, we define the
antisymmetric two-tensor $\delta\Theta$ 
\be
\delta\Theta = \delta\theta^{\alpha \beta}  v_\alpha  v_\beta
\ee
where $v_\alpha$ are one-vectors satisfying $\langle v_\alpha , dx^\beta \rangle
= \delta_\alpha^\beta$ and all products are antisymmetric. Then the contraction
$\langle \delta\Theta , \Tr\D^{2n+3} \rangle$ takes the form
\be
\langle \delta\Theta , \Tr\D^{2n+3} \rangle = (2n+3) \, \delta\theta^{\alpha \beta}
 \sum_{k=0}^n (-1)^k \, \Tr ( D_\alpha \D^k D_\beta \D^{2n+1-k} )
\ee
We observe that (\ref{SWCS}) is the first two terms of the above contraction.

Since we are in $2n+1$ dimensions, all forms with degree higher that $2n+1$ 
vanish identically. Therefore the above contraction, involving $\D^{2n+3}$,
will also vanish. The transformation (\ref{SWCS}), however, differs from that by 
extra terms which generically do not vanish. 
In the special case $2n+1=3$ only, we have
 \be
\delta \Tr \D^3 = \frac{2}{5} \langle \delta\Theta , \Tr\D^5 \rangle = 0
\ee
This is the only case in which the variation of the Chern-Simons action under
a Seiberg-Witten transformation vanishes. For higher dimensions the invariance
of the action is essentially spoiled by noncommutative ordering effects.

\section{Concluding remarks}

We have analyzed the topological properties of the Seiberg-Witten map.
Several issues call for further investigation, however, and we state a few here.

The solution of the Seiberg-Witten map at the point $\theta =0$ was identified
in \cite{OO} in terms of an abelian commutative field strength, for spaces of even
dimension. (The extension to odd dimensions is given in \cite{JPP}.) This, however,
bypasses all the issues on gauge transformations, such as the ones examined
in this paper. It would be desirable to have a solution in terms of gauge potentials 
themselves, which would reveal singularities of the map due to nontrivial
topology. Further, the existing solution holds strictly for U(1) gauge theory. 
The generalization to U(N) gauge theory is an interesting open issue.

On a different front, noncommutative gauge theory describes a kind of fuzzy fluid
\cite{Suss,JPP}. The Seiberg-Witten map is, then, identified as an instance of the
Lagrange to Euler map for fluids. This could be useful to the description of quantum
Hall states in terms of a noncommutative Chern-Simons theory, as proposed by Susskind.
The generalization of the map, however, to the theory describing finite quantum Hall droplets
\cite{APC}, which would be an important ingredient in this description, is not known.

\end{document}